\documentclass{elsart}

\usepackage{natbib}
\usepackage{graphicx}

\usepackage{amssymb}
\usepackage{amsmath}
\def\url#1{{\ttfamily\def\/{/\discretionary{}{}{}}#1}}
\def\bibcode#1{}

\def\Msun{M_\odot}
\begin{document}

\begin{frontmatter}
\title{Using Weak Lensing to find Halo Masses}
\author[address1,address2]{Roland de Putter\thanksref{rdpemail}}, 
\author[address1,address3]{Martin White\thanksref{mwemail}}
\address[address1]{Department of Astronomy, University of California,
Berkeley, CA, 94720}
\address[address2]{Department of Physics, Universiteit Leiden, 2300 RA Leiden,
The Netherlands}
\address[address3]{Department of Physics, University of California,
Berkeley, CA, 94720}
\thanks[rdpemail]{E-mail: R.De.Putter@umail.leidenuniv.nl}
\thanks[mwemail]{E-mail: mwhite@berkeley.edu}

\begin{abstract}

Since the strength of weak gravitational lensing
is proportional to the mass along the line of sight, it might be possible to use lensing data
to find the masses of individual dark matter clusters. Unfortunately, the effect on
the lensing field of other matter along the line of sight is substantial.
We investigate to what extent we can correct for these projection effects
if we have additional information about the most massive halos along the line of sight
from deep optical data.
We conclude that unless we know the masses and positions of halos
down to a very low mass, we can only correct for a small part of the line-of-sight projection,
which makes it very hard to get accurate
mass estimates of individual halos from lensing data.

\end{abstract}

\end{frontmatter}

\section{Introduction} \label{sec:intro}

Weak gravitational lensing is the small distortion, or shearing, of the images of distant galaxies
by the bending of light due to the potentials associated with density fluctuations along the
line of sight. This
means that the lensing
effect on galaxy shapes depends on large scale structure and its evolution over a long period of time, making
weak gravitational
lensing a potentially very useful tool to teach us about cosmology in general and the formation of structures in particular.
Although most recent investigations focus on using the statistical properties of the lensing field to learn about cosmology, lensing
might also be used to determine the masses of individual dark matter halos.
For a recent review of the status of weak lensing research, we refer to \cite{WaerMell} or \cite{Refr}.

Naively, lensing offers a way to find the mass of halos because the
strength of the shear is proportional to the mass along the line of sight. Unfortunately, lensing is caused
by all mass along the line of sight, not just by the particular halo we want to find the mass of. This projection effect
of density fluctuations along the line of sight is known to be considerable \citep{MetzWhi,WhivanW,Pad,HennSp}, which makes
it hard to derive the mass of a halo from the lensing field directly.
However, if we have additional information about the mass along the line of sight, we may be able to correct for
its effect, at least partly (we refer to \citealt{Dodel} for a more statistical approach to correcting for line-of-sight
projection). One way of getting this information is to use deep optical data, that we need anyway
to detect background source galaxies,
to find halos by applying modern cluster finding methods (see for example \citealt{GladYee} on RCS and
\citealt{Bahcall} on SDSS). More specifically, halos can be located by
looking for groups of galaxies, using the fact that galaxies tend to trace the dark matter distribution.
If we then correct for
the lensing effect of these line-of-sight halos, we might be able to retrieve the lensing signal caused purely by the 
particular halo of interest, which is a measure of its mass. Knowledge of halo masses would be very useful
from a cosmological point of view because it can be used to construct
the mass function, which would help us to provide better constraints on cosmological parameters such as the equation
of state of the dark
energy \citep{Mohr,Mohretal}.
In this paper we will investigate to what extent we can correct for line-of-sight projection effects in the lensing field
and we will show that it is unlikely that the lensing field can be used to determine individual halo masses with a
higher accuracy than the 10\% level.
In section \ref{sec:model}, we will discuss the theory and the practical aspects of our analysis.
The results are discussed in section \ref{sec:results}.

\section{Model} \label{sec:model}

We will analyze the method described above by considering lensing
maps based on simulations of structure formation.
We first discuss these simulations and how they are used to calculate
lensing fields in section \ref{subsec:sim}.
These fields will play the role of the observed data. To find out if
we can correct for the lensing effect of line-of-sight halos and use the
lensing field to find halo masses, we will use information about the
positions and masses of these halos together with a model for their density profiles
to calculate their contributions to the measured field. This halo model will be discussed in
section \ref{subsec:halo}. Readers only interested in the results are referred to
section \ref{sec:results}. 

\subsection{N-body simulations} \label{subsec:sim}

The basis for the lensing maps is a high resolution N-body simulation of
structure formation in a $\Lambda$CDM universe which was run using the
TreePM code described in \cite{TreePM}.
The model (specifically Model 1 of \citealt{YanWhiCoi})
assumed $\Omega_{\rm mat}=0.3$, $\Omega_\Lambda=0.7$, $h=0.7$, a
scale-invariant primordial spectrum and $\sigma_8=1$.
The simulation used $512^3$ dark matter particles of equal mass
$1.7\times10^{10}\,h^{-1} \Msun$ in a cubical box
of fixed comoving size
$300\,h^{-1}$Mpc on a side with periodic boundary conditions.
The softening was of a spline form with Plummer equivalent smoothing
$20\,h^{-1}$kpc, fixed in comoving coordinates.
The simulation was started at a redshift $z=60$ when density fluctuations on 
the relevant scales were still in the linear regime. Between $z=2$ and $z=0$
the particle distribution was dumped every $100\,h^{-1}$Mpc.

For each output we produced a halo catalog by running a
``friends-of-friends'' (FoF) group finder (e.g.~\citealt{DEFW}) with
a linking length $b=0.15$ in units of the mean inter-particle spacing.
This procedure partitions the particles into equivalence classes, by 
linking
together all particle pairs separated by less than a distance $b$. This means that
FoF halos are bounded by a surface of density roughly $140$ times
the background density. More than half of all the particles are unclustered and
are assigned to the ``no group'' category. Note
that since simulated dark matter clusters do not have clear boundaries,
our definition (and any other practical definition for that matter) excludes
a small part of each cluster from the FoF cluster.
The halo mass is estimated as the sum of the masses of the particles in 
the
FoF halo, times a small correction factor (typically 10\%) which 
provides
the best fit to the Sheth-Tormen mass function
\citep{SheTor}.
The resulting catalog contains the angular position in the sky, the redshift
and the mass for each halo.
Since the typical source distance is a lot bigger than the box size
(for a
source redshift $z_s=1$ the distance is about $2300\,h^{-1}$Mpc),
we place a number of boxes in a row to get the complete matter distribution between
$z=0$ and $z=z_s$ within a field of view of $3^\circ$.
The origin of each box is chosen at random in order to prevent tracing
through the same structure more than once.

As a measure for the lensing
effect, we focus on the convergence $\kappa$ because this is an easy quantity
to work with. While it is possible to use a full ray tracing algorithm
\citep{JSW,ValeWhite} to compute $\kappa$ from the simulations, we used a
simpler approach, based on the Born approximation, which should be more
than adequate for our purposes \citep{HuWhite}.
The Born approximation gives \citep{WaerMell}:
\begin{equation}
  \kappa \simeq \frac{3}{2}H_0^2\Omega_{\rm mat} \int_{0}^{\chi_s} d\chi
   \frac{\chi(\chi_s-\chi)}{\chi_s}\frac{\delta}{a}
\end{equation}
where $\delta$ is the overdensity, $a$ is the scale-factor and $\chi$ is the
comoving distance. The integral is along a straight line between observer
and source. We thus calculate maps of $1024\times1024$ pixels corresponding
to a $3^\circ\times3^\circ$ field of view. For simplicity we use a fixed
source redshift $z_s = 1.05$.

\subsection{The Halo Model} \label{subsec:halo}

In order to find the mass of a particular halo we wish to use the lensing field around
it and correct
for the expected effect of other observed halos along the line of sight. 
To calculate this correction we must assume a certain density
profile for each line-of-sight halo.
The density profile we use in our model is the NFW distribution
(Navarro, Frenk \& White 1997,1996,1995).
Previous simulations of structure formation have shown that most dark matter halos
roughly follow this profile, which is given by:

\begin{equation}
\rho(r) = 
\frac{\delta_{c}\rho_{c}}{\left(r/r_{s}\right)
\left(1+r/r_{s}\right)^{2}},
\label{rhonfw}
\end{equation}
where $r$ is the distance to the halo center and $\rho_{c} = 3H^2(z)/8\pi G$
is the critical density
at the redshift $z$ of the halo.
The scale radius $r_{s} = r_{200}/c$ is a characteristic radius of 
the cluster, 
$c$ is a dimensionless number known as the concentration parameter, and 
\begin{equation}
\delta_{c}= \frac{200}{3}\frac{c^{3}}{\ln(1+c)-c/(1+c)}
\end{equation}
is a characteristic overdensity for the halo.  
The virial radius, $r_{200}$, is defined as 
the radius inside which the mass density of the halo is equal to
$200 \rho_c$.
 
We assume that for $r>r_{200}$ $\rho(r) = \overline{\rho}(z)$ where
$\overline{\rho}(z)$ is the mean mass 
density of the universe at redshift $z$. In other words,
the halo ends at $r_{200}$. Previous analyses of simulations of structure
formation have shown that $c$ is redshift and mass dependent.
We use the concentration relation from \cite{Bullock}:
\begin{equation}
c(M,z) = \frac{9}{1+z}\left[\frac{M}{M_{\star}(z)}\right]^{-0.13} \, .
\label{c}
\end{equation}
Here $M_{\star}$ is defined by $\sigma(M_{\star},z) = \sigma(M_{\star},0)g(z) = 1.686$, where
$\sigma(M,z)$ is the standard deviation in the relative density of a region
of volume $V = M/\overline{\rho}(z)$ at redshift $z$ and $g$ is
the growth factor.
In our model, we use Eq (\ref{c}) for
$c(M,z)$ where we take the transfer function from \cite{EisHu} to calculate
$\sigma(M,0)$ and the fitting function from \cite{Wang} for $g(z)$. This gives values of
$M_{\star}(z)$ ranging from $1.5\times10^{12}\,h^{-1}\Msun$ to $1.5\times10^{13}\,h^{-1}\Msun$
from $z = 1$ to $z = 0$.
Given the above equations, we can calculate the
density as a function of distance to the halo center if we know $M$
and $z$.

\section{Results} \label{sec:results}

Among other things, to what extent we can correct for projection effects depends on
how much of the convergence
is caused by the set of halos that we are able to identify in an optical survey.
Since only the more massive clusters are likely to be found
by looking for groups of galaxies, the
contributions to the convergence from 
smaller clusters and matter that is not part of any cluster can not be corrected
for at all and therefore needs to be sufficiently small
for our method to work (\citealt{MetzWhi} also adressed this question to some extent).
To give a rough idea of the contributions to the field
of halos in different mass ranges, we first analyze the convergence power spectrum
in section \ref{subsec:mass}. In section \ref{subsec:findmass} we then turn our attention
to the tangential shear
$\gamma_{\rm T}$ around (the most massive) halos and analyze the effect on this quantity of
other matter along the line of sight. 
Next, we try to compensate for the effect of ``known'' halos by subtracting their effect
on the shear using the halo model described above. Finally, in section \ref{subsec:galcounts}
we incorporate the uncertainty in the masses of the line-of-sight halos that arises from the
fact that their masses will likely be estimated from cluster richness measurements.

\subsection{Power spectrum as a function of halo mass} \label{subsec:mass}

\begin{figure}[h]
\begin{center}
{\includegraphics*[height=6.5cm,width=6.5cm]{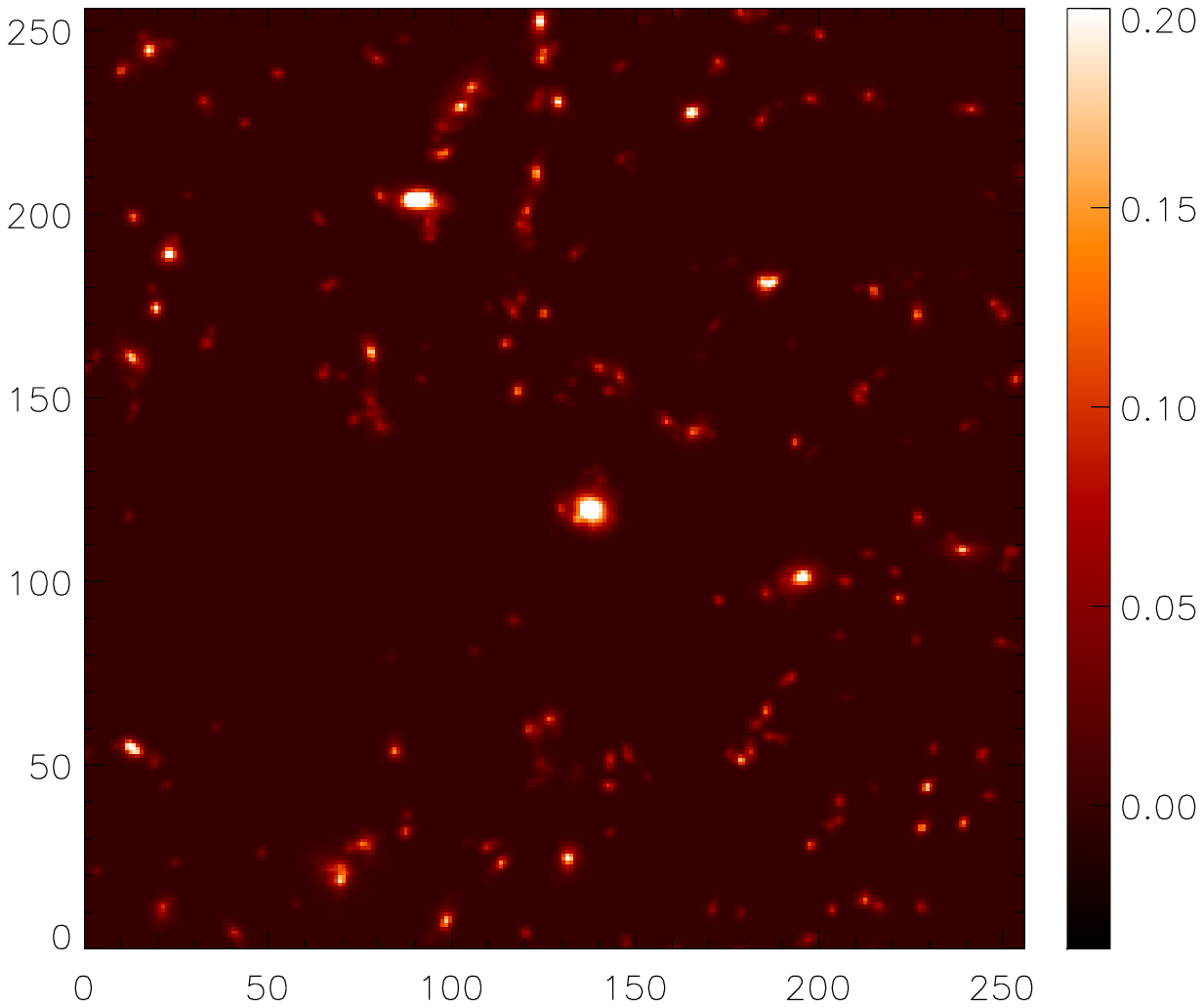}
 \includegraphics*[height=6.5cm,width=6.5cm]{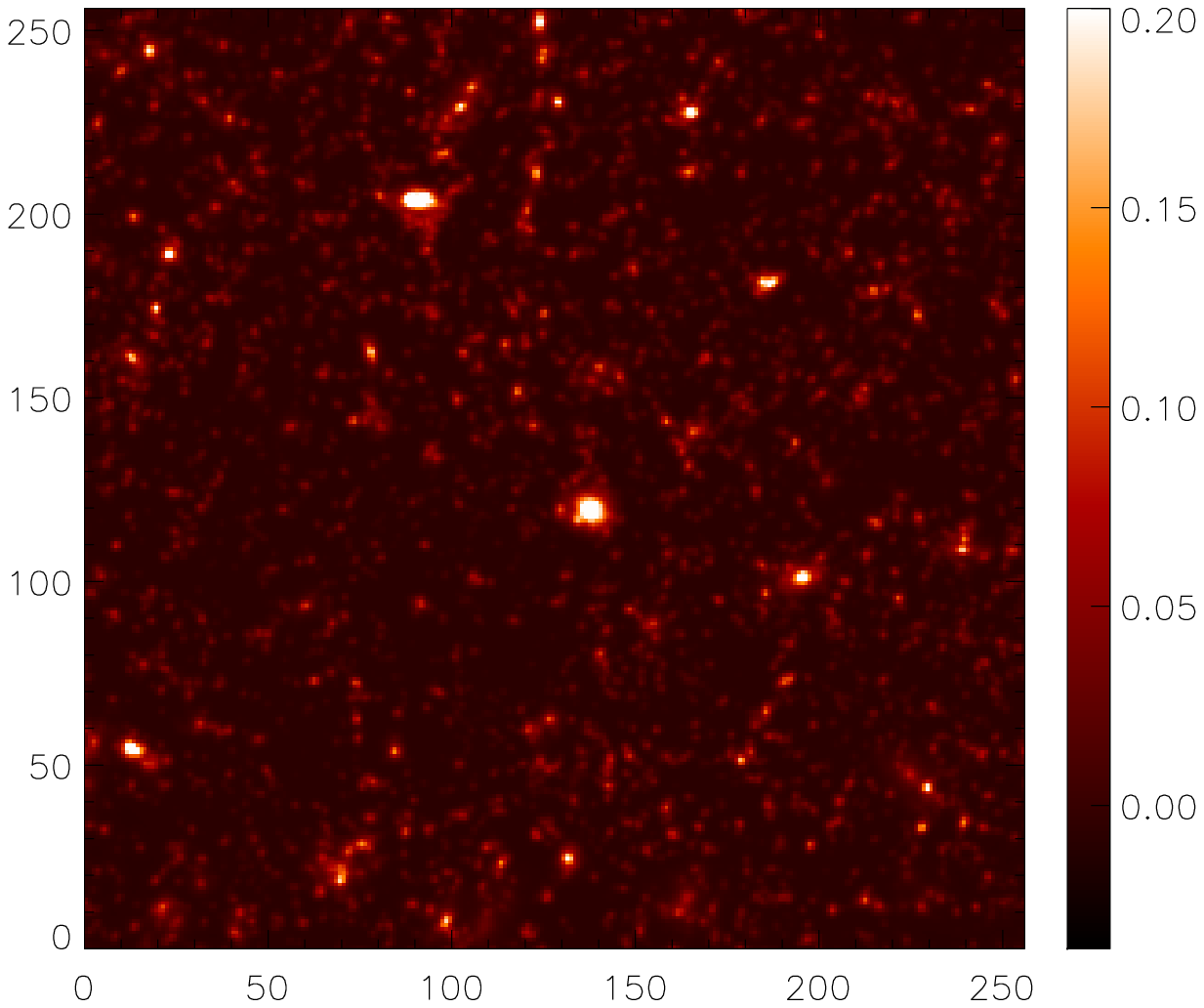}}
{\includegraphics*[height=6.5cm,width=6.5cm]{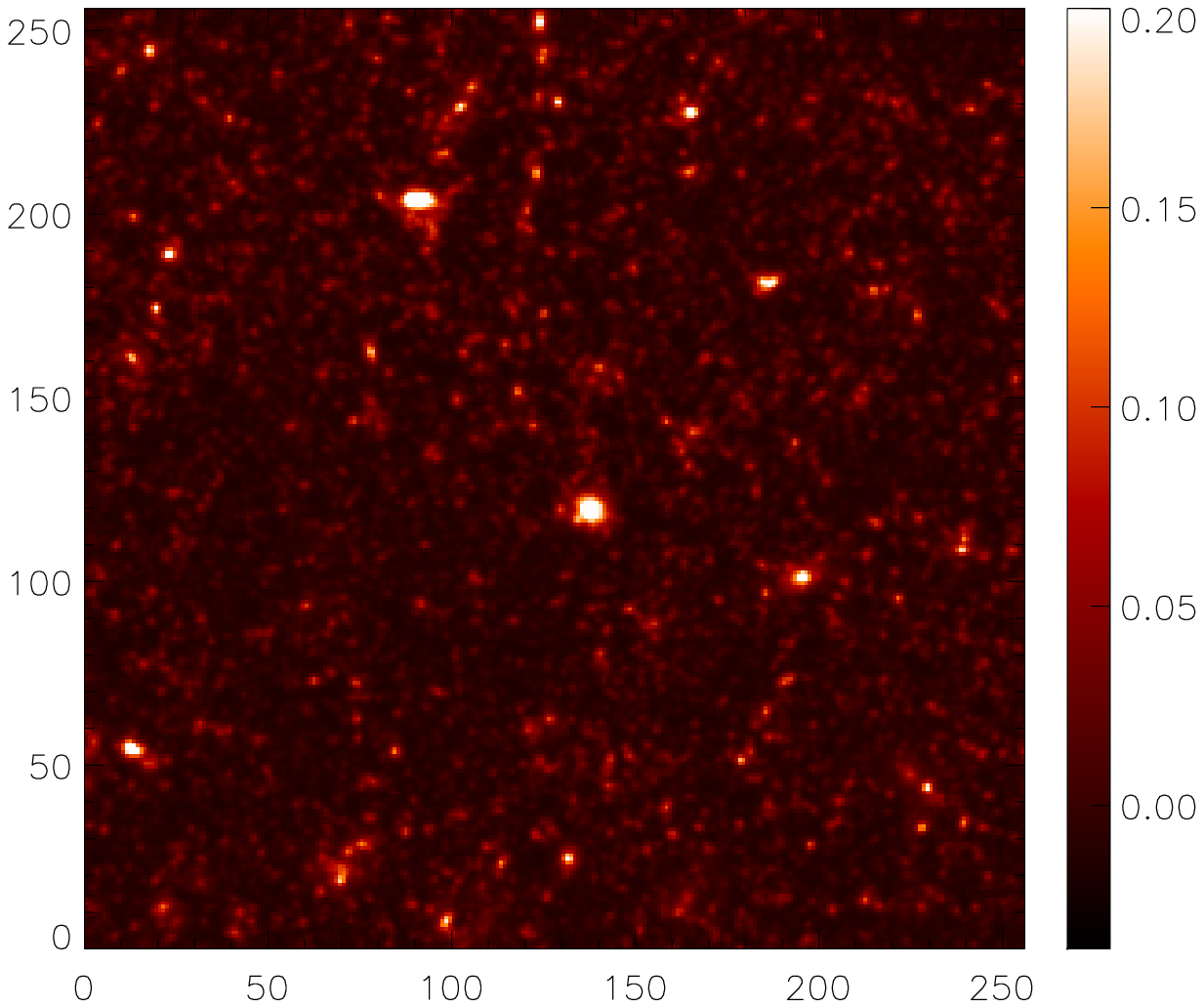}
\includegraphics*[height=6.5cm,width=6.5cm]{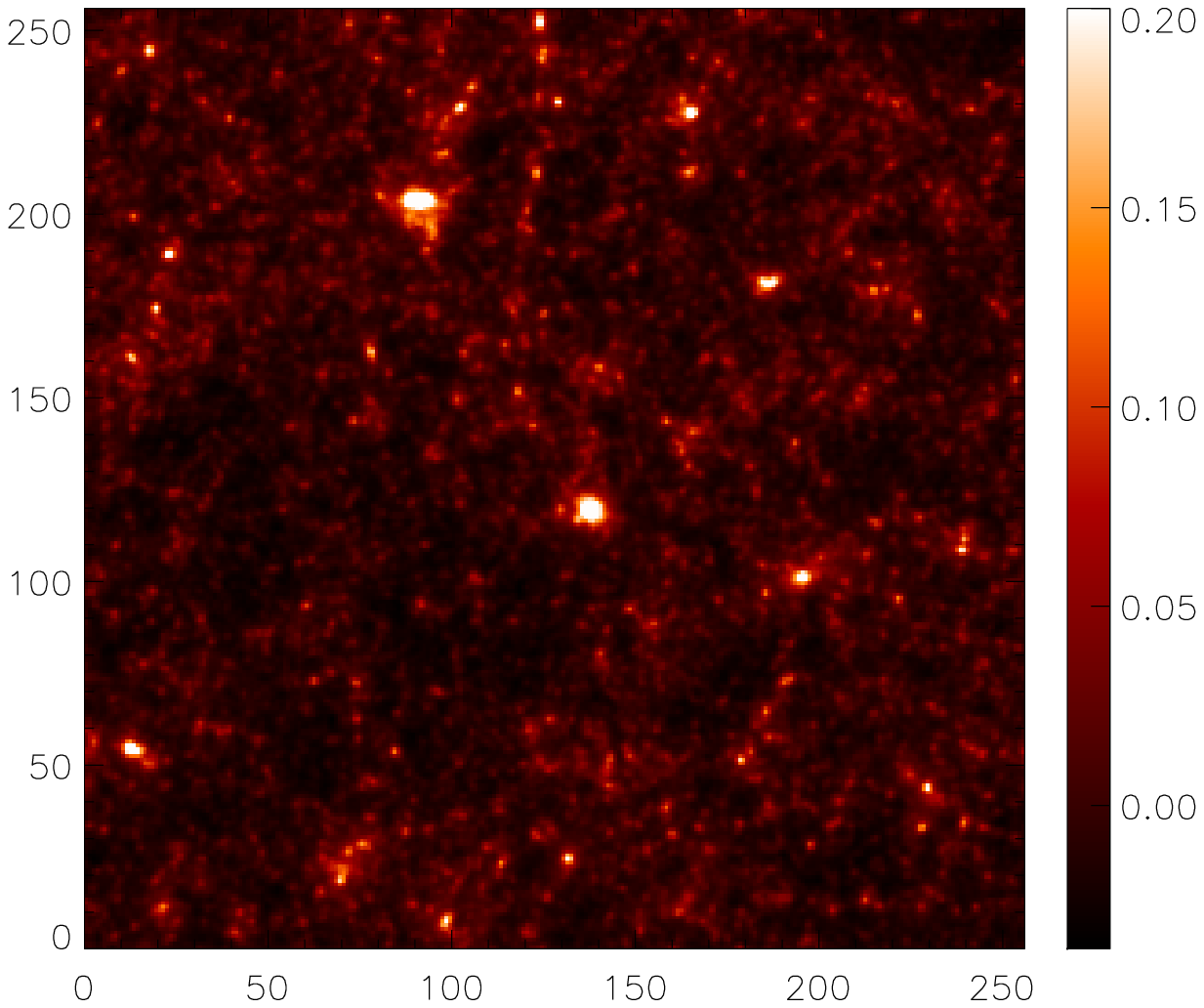}}
\end{center}
\caption{Convergence maps ($3^\circ\times 3^\circ$)
from different selections of particles.
From upper left to lower right: only particles in halos more
massive than
$M=10^{14}$, $10^{13}$ and
$10^{12}\,h^{-1}\Msun$ and finally, all particles.
The maps have been smoothed with a Gaussian with FWHM$=1'$ and
the scale is between -0.036 and 0.2.
}
\label{fig:Maps}
\end{figure}

\begin{figure}[h]
\begin{center}
\includegraphics[width=12cm]{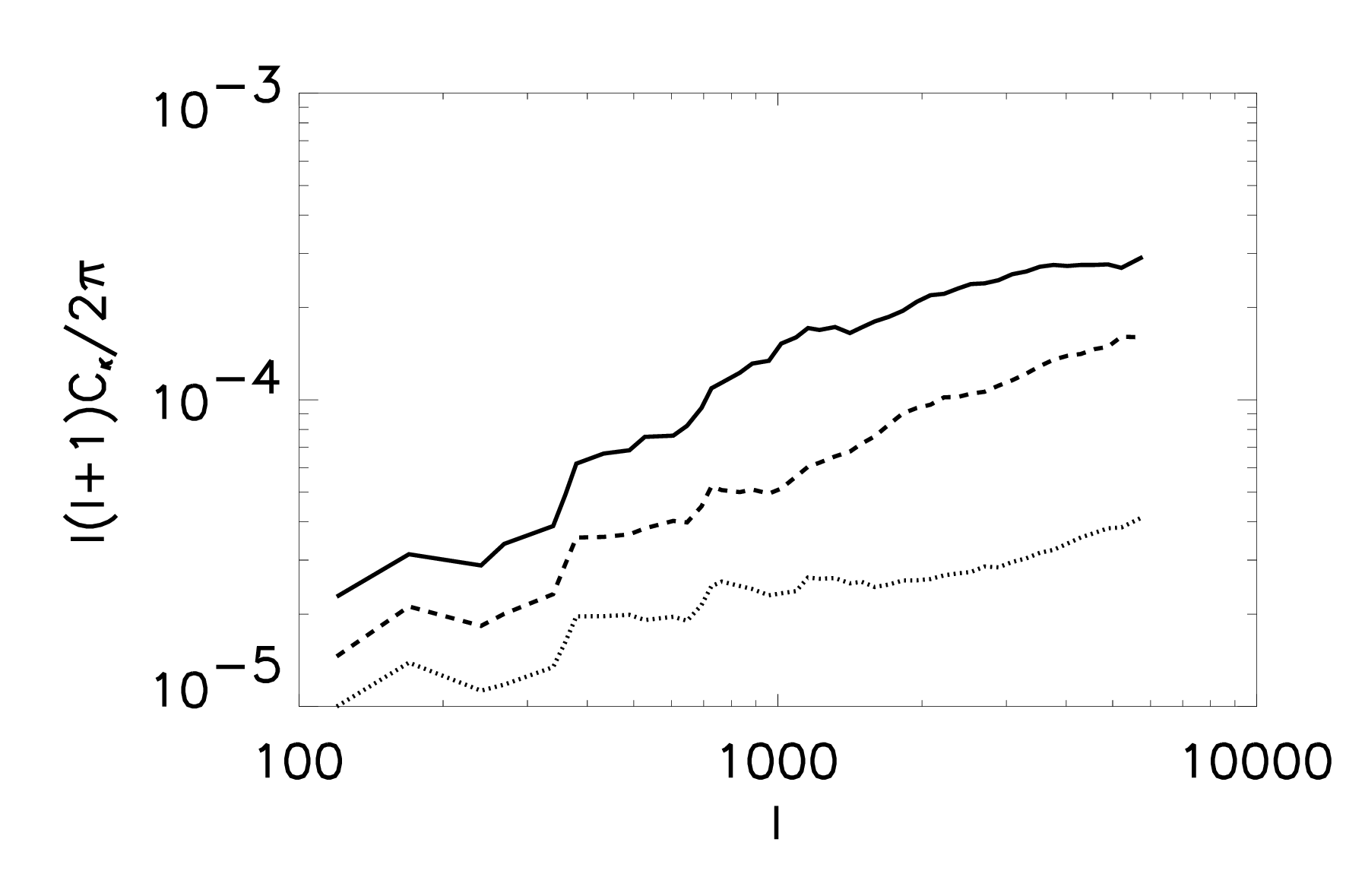}
\end{center}
\caption{Convergence power spectra resulting from simulations of
structure formation. We show the spectrum resulting from all structure
except halos with mass above masscuts of
$10^{15}\,h^{-1}\Msun$ (solid),
$10^{14}\,h^{-1}\Msun$ (dashed) and
$10^{13}\,h^{-1}\Msun$ (dotted). On average, the
power from halos less massive
than $10^{14}\,h^{-1}\Msun$ is about a factor 2 lower than the total power. 
}
\label{fig:Power}
\end{figure}

Using the simulation data and the group information of each particle, we construct convergence maps for
different selections of particles. Structure that is left out this way is replaced by a uniform matter
density such that all maps have the same average convergence.
In Fig. \ref{fig:Maps} we show the contributions to the convergence field
from halos with masses above
$10^{14}$, $10^{13}$ and $10^{12}\,h^{-1}\Msun$.
In Fig. \ref{fig:Power} we show the power spectra of 
maps where the effect of halos more massive than certain mass cuts has been taken out.
The cuts are chosen to be $10^{15}, 10^{14}$ and $10^{13}\,h^{-1}\Msun$ in order to be able
to compare the results to the analytical results of \cite{Coorayetal}. Fig. \ref{fig:Power} shows
that the power spectrum coming from matter in halos with mass below $10^{14}\,h^{-1}\Msun$ and ungrouped
matter
is about half of the total, which agrees nicely with their predictions. The power from halos less massive
than $10^{13}\,h^{-1}\Msun$ and ungrouped matter is roughly 10\% - 20\%. Since the selection
of line-of-sight halos we will be able to find in optical data will be determined by a (redshift
dependent) minimum mass, very roughly corresponding to the mass cuts in Fig. \ref{fig:Power}, we
already see that a
significant part (in terms of the
power spectrum) of the projection effect will be difficult to correct for.

\subsection{Deprojecting known halos} \label{subsec:findmass}

\begin{figure}[h]
\begin{center}
\includegraphics[width=12cm]{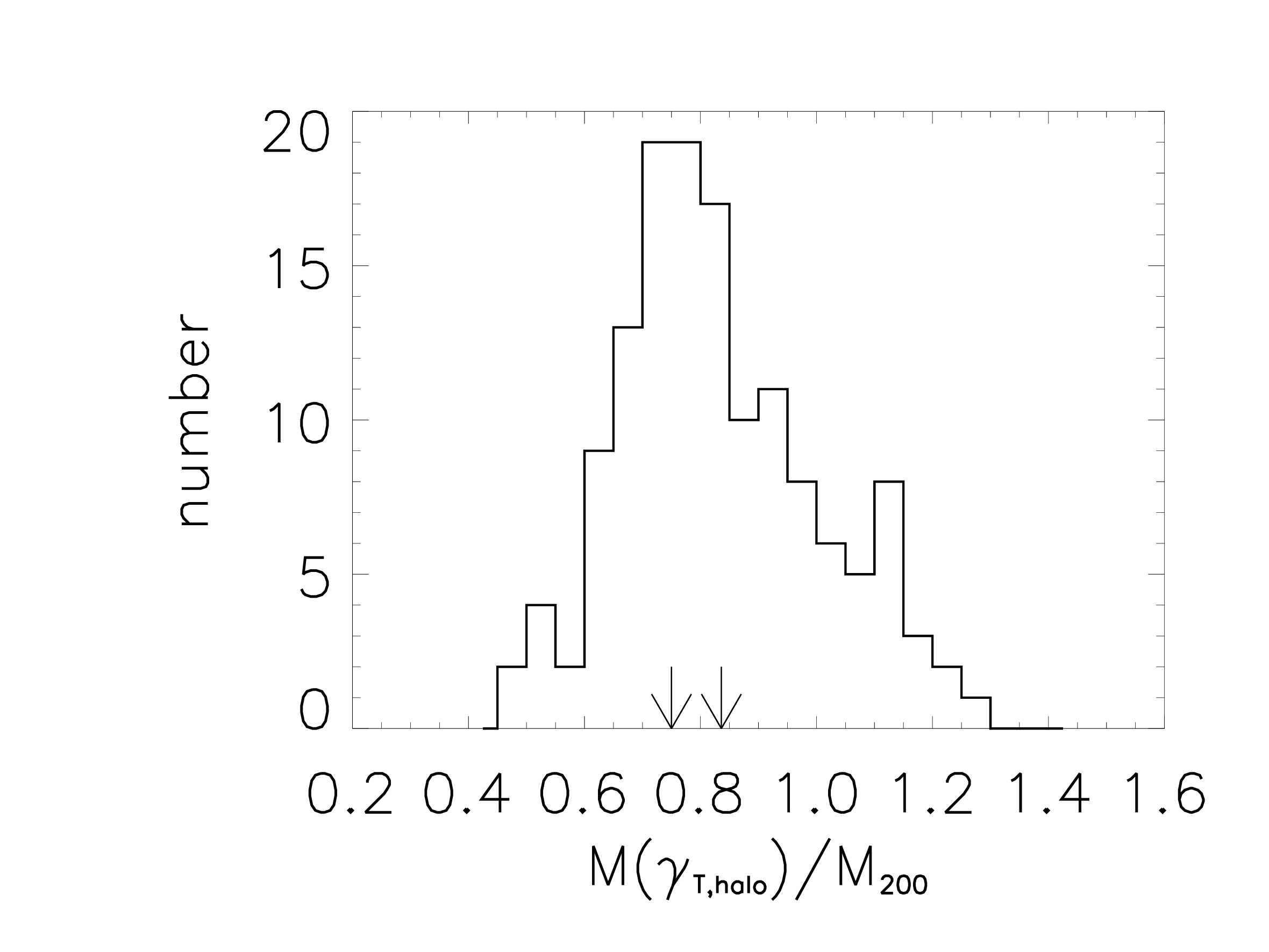}
\end{center}
\caption{Scatter in mass estimates from $\gamma_{\rm T,halo}$ with respect to the virial mass $M_{200}$ for
the set
of 141 target halos defined in the text. $\gamma_{\rm T,halo}$ is the shear
at $r=r_{200}/2$ purely caused by the halo of interest, so line-of-sight projection has not yet been taken into account.
The arrows indicate the median (left) and the mean (right).
The histogram that even if we are able to correct for the projection effect of all line-of-sight structure perfectly,
we still have to deal with the fact that the enclosed mass within $r$ may differ from $M_{200}$ and
the fact that $\gamma_{\rm T,halo}$ does not exactly correspond to the enclosed mass if $\kappa(r)\not=0$.
}
\label{fig:mvir}
\end{figure}

\begin{figure}[h]
\begin{center}
\includegraphics[width=12cm]{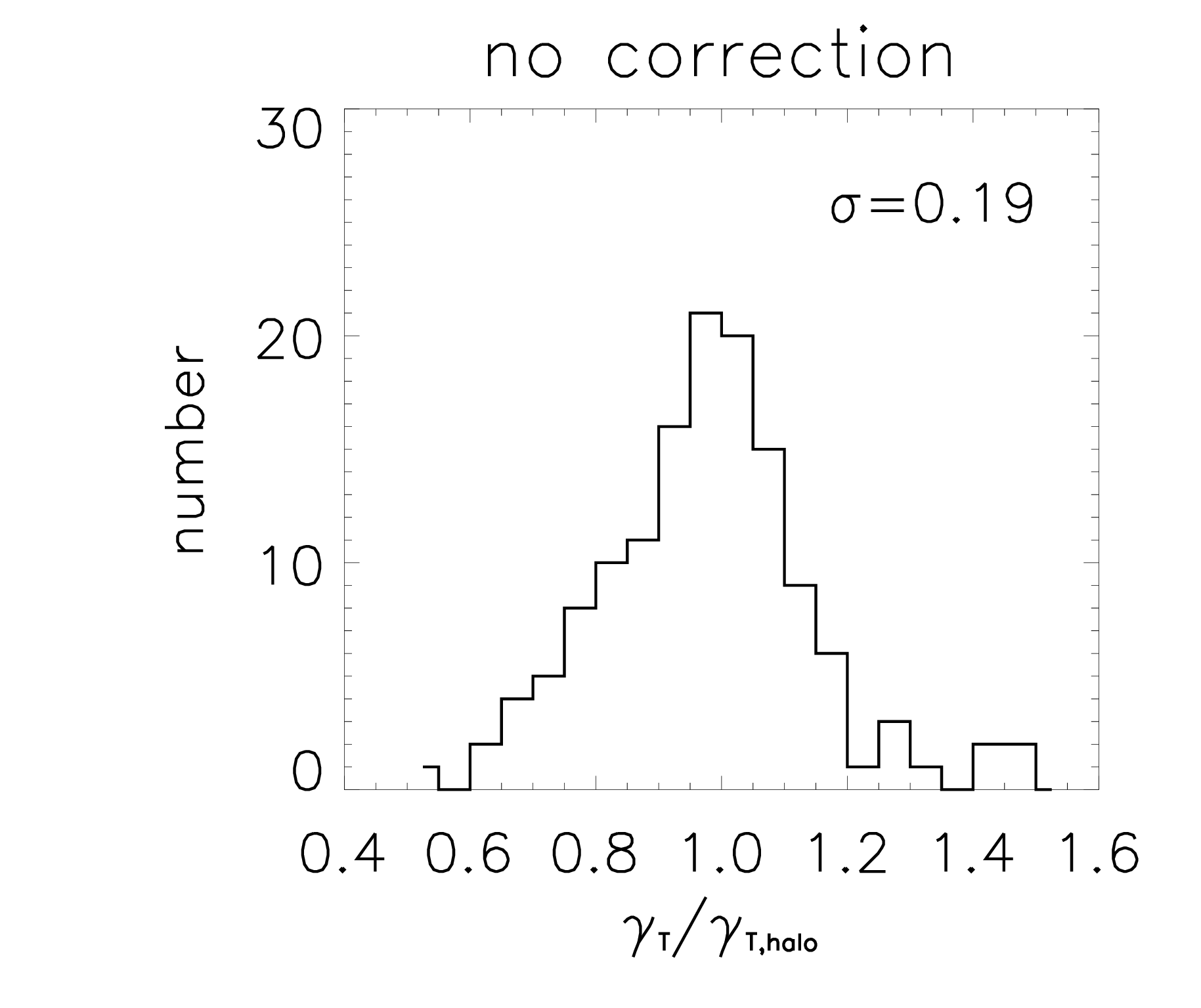}
\end{center}
\caption{Scatter in mass estimates based on the transverse shear for
a sample of 141 massive halos defined in the text.
$\gamma_{\rm T}$
is the average shear at radius $r_{200}/2$. Projection effects of line-of-sight structure cause a
large scatter in mass estimates derived directly from the measured shear.
}
\label{fig:all}
\end{figure}

In this and the following section we investigate to what extent we can eliminate line of sight projection
if we know
the masses and positions of (some of) the line-of-sight halos. As a measure for the mass enclosed within
a transverse radius $r$ of the halo center we
use $\gamma_{\rm T}(r)$, the {\rm average} tangential shear at $r$, because unlike the convergence this is a
quantity that can be observed. It
is related to the convergence by
\begin{equation}
\gamma_{\rm T}(r) = \kappa(<r) - \kappa(r)
\end{equation}
(see for example \citealt{Mandel}). Here $\kappa(<r)$ and $\kappa(r)$ are
the average convergences
within the area with radius $r$ and at $r$ respectively. The mass of a certain halo within $r$
relates to the tangential shear caused only by that halo $\gamma_{\rm T,halo}$
through:
\begin{equation}
M_{\rm encl}(r) \propto \frac{c^2}{4\pi G}\frac{\chi_s}{(1+z)\chi(\chi_s-\chi)}A_{\rm halo} \gamma_{\rm T,halo}(r),
\label{Mgamma}
\end{equation}
where $A_{\rm halo}$ is the projected surface area of the region within $r$ in units of comoving distance squared and the
relation is an equality if $r$ is big enough for $\kappa(r)$ to be zero since in that case
$\gamma_{\rm T,halo}(r)=\kappa_{\rm halo}(<r)$.
If $\kappa(r)>0$, we need to multiply
$\gamma_{\rm T,halo}$ by a correction factor before we can relate it to the mass by the above equation.
If the halo has an NFW profile with $c\approx4$, this factor will be about $1.6$ at a radius $r=r_{200}/2$.
We calculate the shear for a set of target halos defined by $M>3\times10^{14}\,h^{-1}\Msun$ and comoving
distance between $500$ and $1900\,h^{-1}$Mpc, so that the lensing kernel is at
least about half of its maximum. 
Using 10 realisations this gives us a set of 141 halos.

\begin{figure}[h]
\begin{center}
\includegraphics[width=14cm]{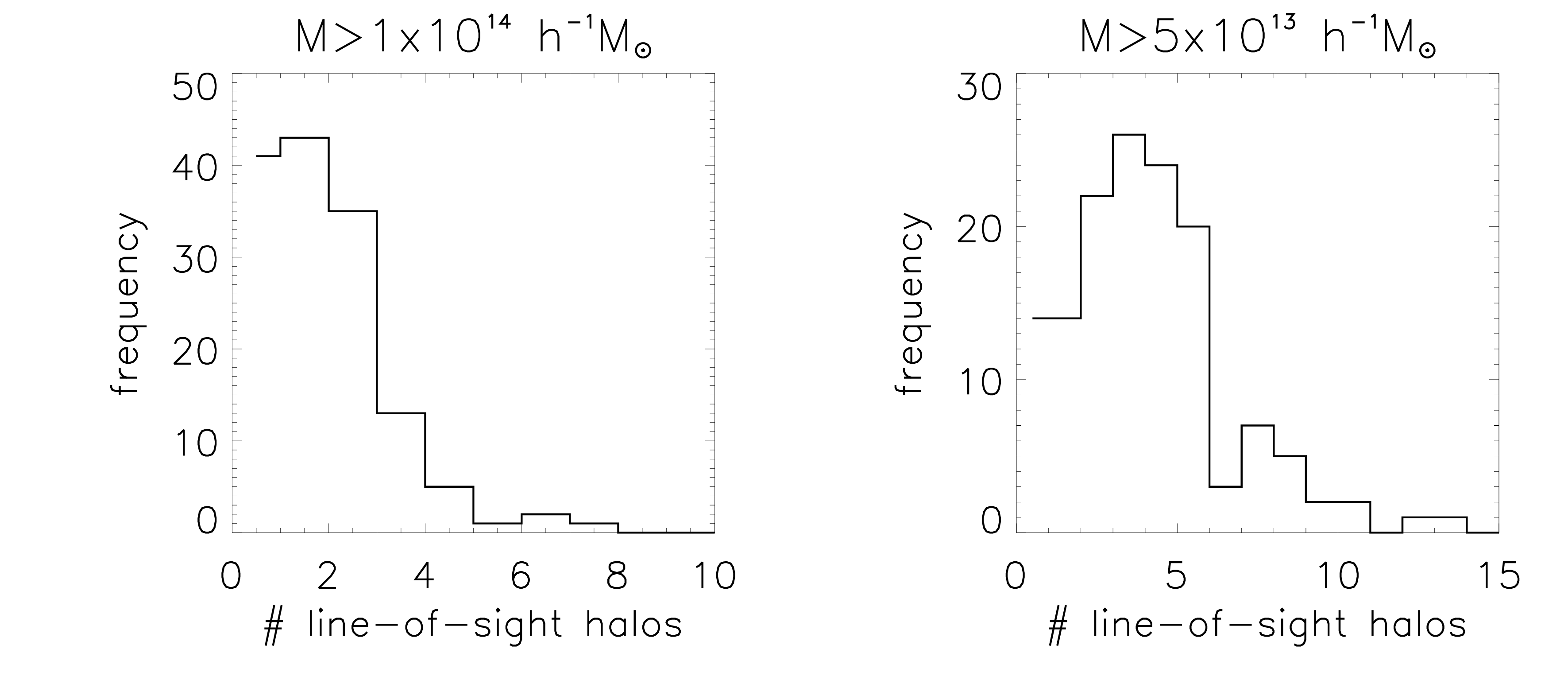}
\end{center}
\caption{Distribution of the number of halos above mass cuts of $10^{14}\,h^{-1}\Msun$ and $5\times10^{13}$
that lie along the line of sight of a certain halo. The number of halos along the line-of-sight was calculated for a set
of 141 massive halos (see text) and a halo is considered
to lie on the line of sight of one of those halos if their projected separation is
less than the sum of their virial radii. We will assume that line-of-sight halos more massive than the above mass
thresholds can be identified in optical surveys.
The histograms show that there is only a small number of these halos along the line of sight of each target halo
and therefore one might expect that they account for only a small part of the line-of-sight projection.
}
\label{fig:los}
\end{figure}

The transverse radius $r$
the shear is evaluated at is taken to be $r_{200}/2$. The disadvantage of this choice of $r$
is that this way we do not get information about the mass of the part of the halo that lies outside of $r_{200}/2$. However, for
larger radii, a lot of pixels with low signal will be included and the signal to noise ratio consequently will go down.
We checked that $r=r_{200}/2$ gives shear measurements with less noise from projection than larger values of $r$.
In Fig. \ref{fig:mvir} we show the relation
between the virial mass and the mass enclosed within $r=r_{200}/2$ as calculated from $\gamma_{\rm T,halo}(r)$, using Eq (\ref{Mgamma})
with a correction factor of $1.6$. $\gamma_{\rm T,halo}$
is obtained from the lensing field with just the halos more massive than $3\times10^{14}\,h^{-1}\Msun$ in it. Since there
are only a few halos more massive than $3\times10^{14}\,h^{-1}\Msun$ in each map, this practically means
that for each target halo, $\gamma_{\rm T,halo}$ is the shear purely caused by that halo.
The scatter is due to the fact that the ratio of $\kappa(r)$ over $\gamma_{\rm T,halo}$ is not always exactly 1.6 and, more importantly,
the fact that the mass enclosed does not in general equal $M_{200}$ (see for example \citealt{MetzWhi} for
a more detailed discussion of this issue). The enclosed mass for example has
a large dependence on the orientation of the halo in case it is elliptical. In the rest of this paper we will ignore the problem of
calculating the (virial) mass in case $\gamma_{\rm T,halo}$ is known and instead focus on the problem of how to find $\gamma_{\rm T,halo}$
in the first place.

\begin{figure}[h]
\begin{center}
\includegraphics[width=14cm]{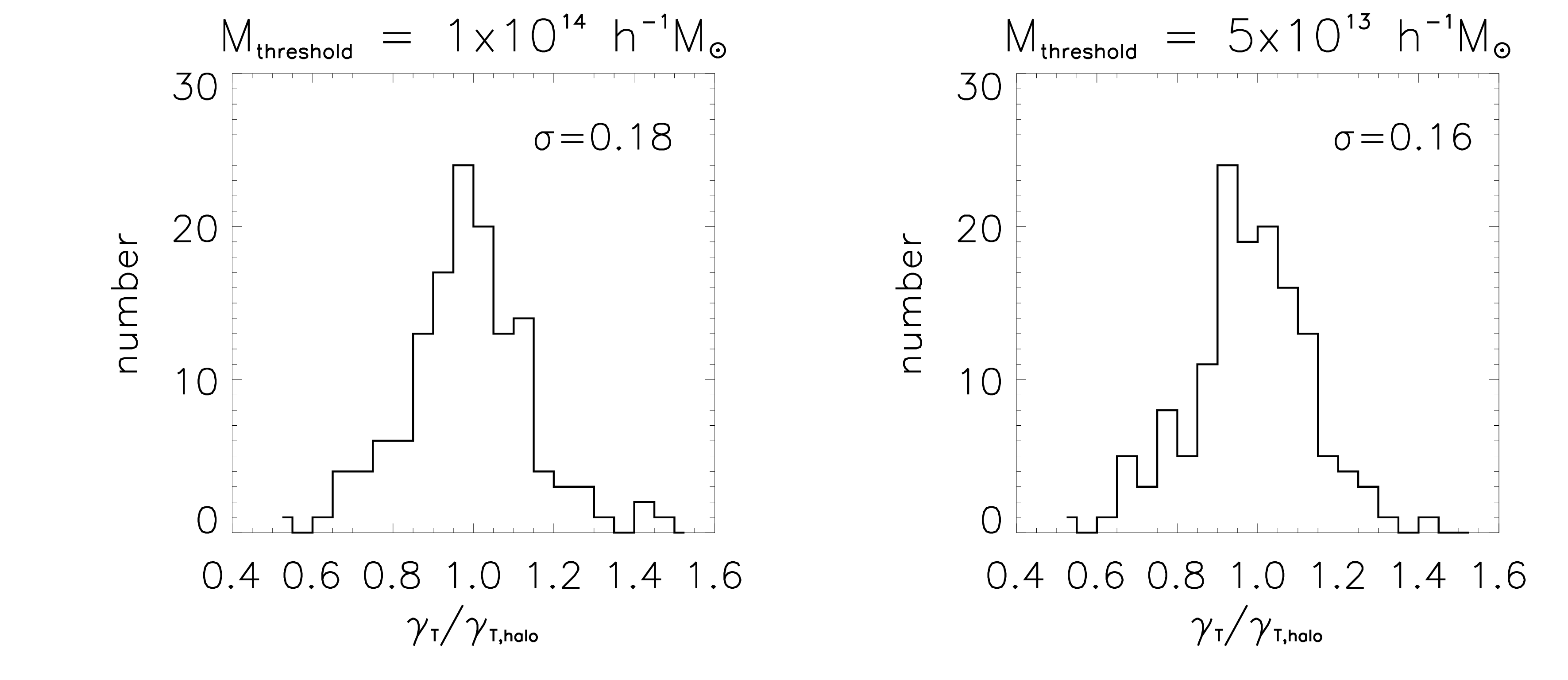}
\end{center}
\caption{Scatter in mass estimates based on the transverse shear for
a sample of 141 massive halos defined in the text.
$\gamma_{\rm T}$
is the average shear at radius $r_{200}/2$ after correcting for the projection effect of line-of-sight halos more massive
than $M_{\rm threshold}$ and $\gamma_{\rm T,halo}$ is the shear purely caused by the halo of interest.
Since we completely correct for the effect of halos above the thresholds, these results show that most
of the scatter is caused by halos below the mass thresholds and by unclustered matter.
}
\label{fig:hisreal}
\end{figure}

In Fig. \ref{fig:all} we show how much
the ``measured'' shear $\gamma_{\rm T}$, caused by all structure along the line of sight, deviates from the shear that is
caused by just the halo of interest $\gamma_{\rm T,halo}$.
In light of the discussion in the previous paragraph, Fig. \ref{fig:all} shows that projection effects will cause
large errors in mass estimates obtained
from the uncorrected lensing field.
Fig. \ref{fig:los} shows the statistics of how many
halos more massive than mass thresholds of $1\times10^{14}$ and $5\times10^{13}\,h^{-1}\Msun$
there are along the line of sight of each of the target halos. We choose these particular thresholds
because virtually all halos more massive than these thresholds can be detected from
deep optical surveys \citep{GladYee} while for lower thresholds this will become very difficult.
We correct for the lensing effect by these line-of-sight halos by
using the actual matter distribution (from the N-body simulations) of each line-of-sight halo to calculate its
contribution to the lensing field and subsequently subtract this from the ``measured'' total field.
We also include in this correction all structure within $3r_{200}$ of the halo center that
is not part of the halo
by our FoF halo definition described in section \ref{subsec:sim} because we do not want to exclude the lensing
effect of structure
that can be considered part of the halo but is not part of the FoF halo.
In Fig. \ref{fig:hisreal} we show that the scatter only gets a little smaller when we correct for the lensing field caused
by halos above the thresholds.
These plots provide us with an upper limit on how well we
can correct for halos above the given mass thresholds.
In reality, we do not know the exact profiles of the line-of-sight halos,
which means we have to resort to a halo model (see section \ref{subsec:halo}) to correct
for their effect.

\begin{figure}[h]
\begin{center}
{\includegraphics*[height=6.5cm,width=6.5cm]{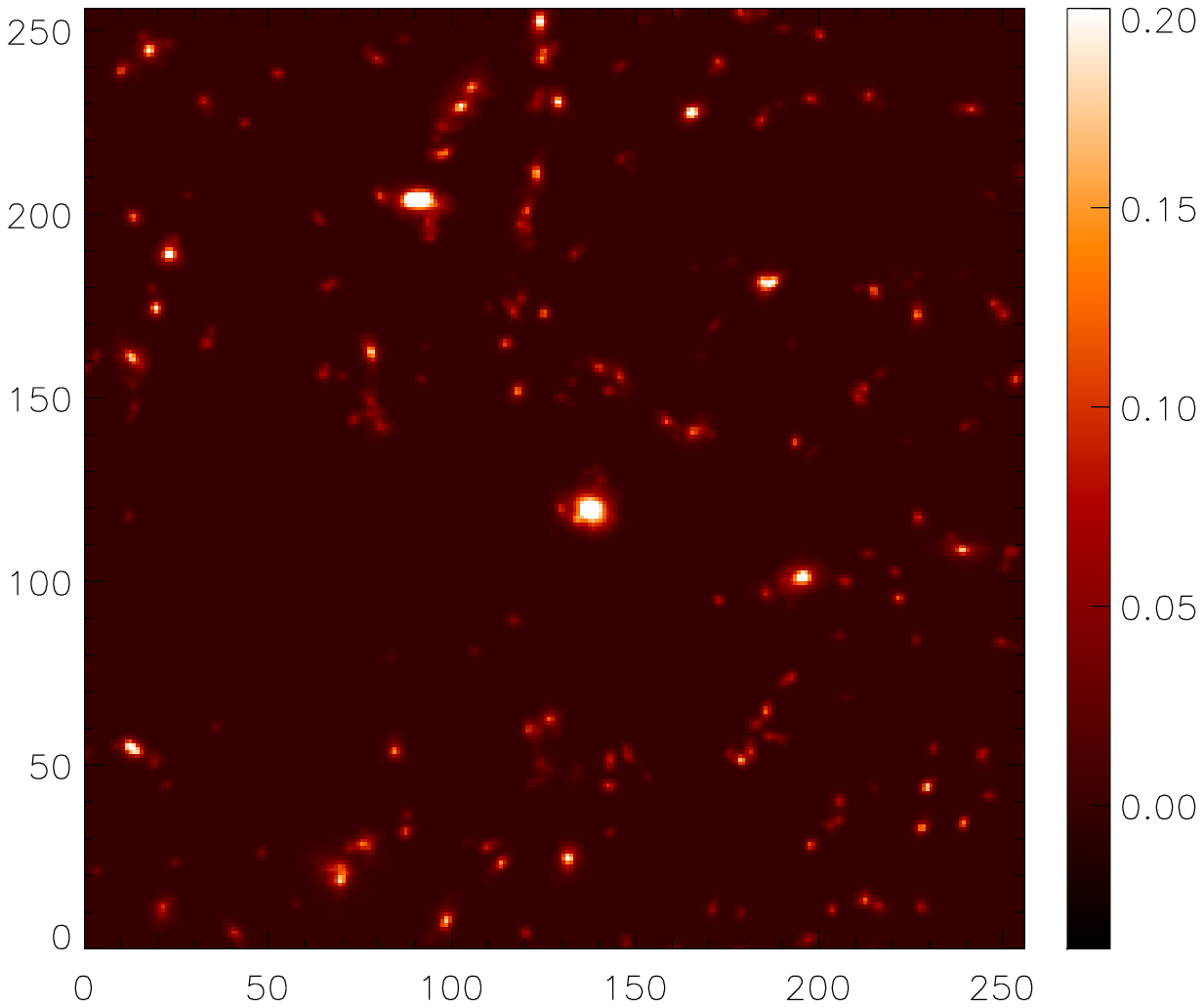}
 \includegraphics*[height=6.5cm,width=6.5cm]{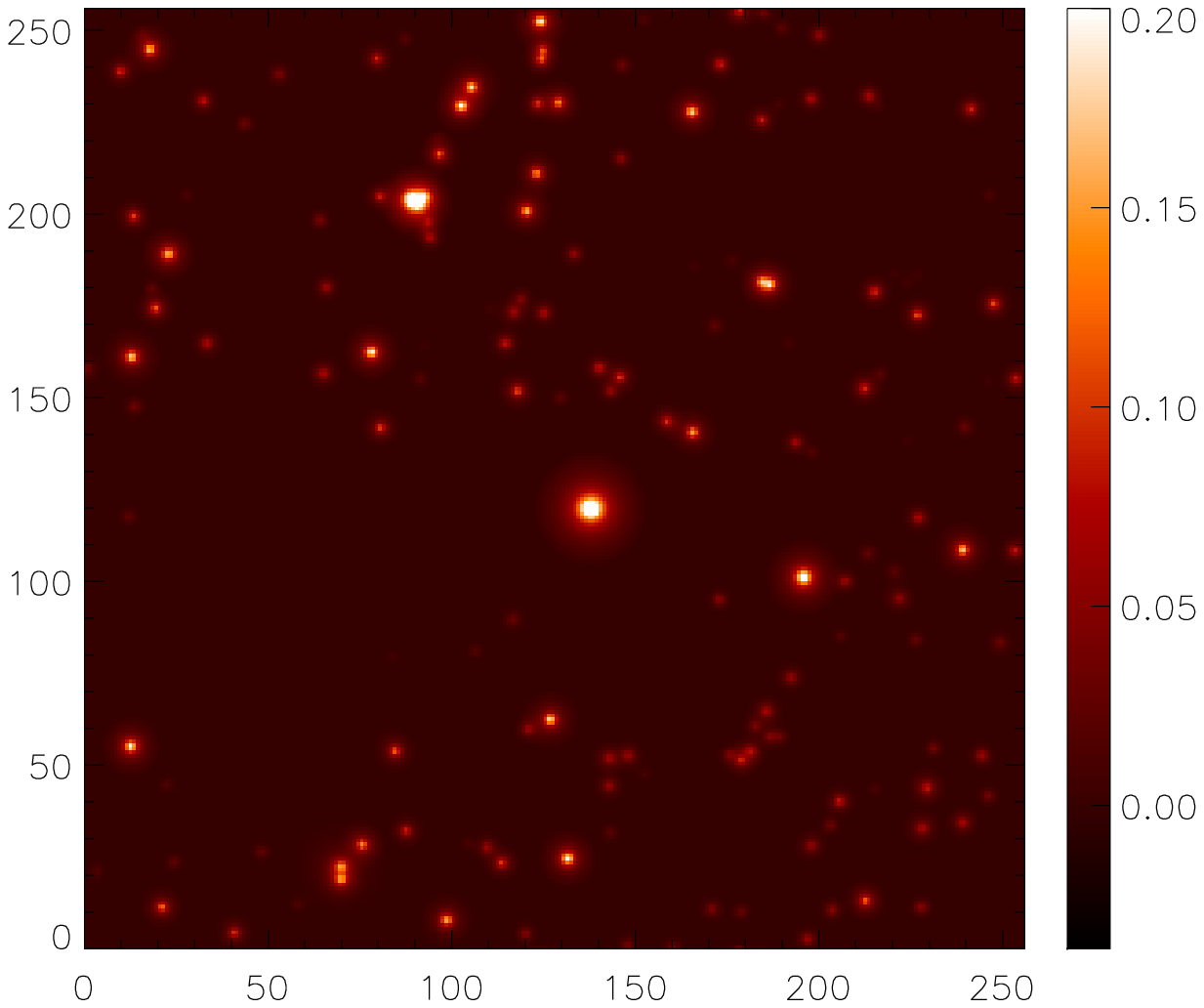}}
\end{center}
\caption{Convergence maps that only take into account halos more massive than 
$10^{14}\,h^{-1}\Msun$. On the left
we use the simulated matter distribution. On the right,
we use the halo catalog together with our
NFW halo model. We use maps of this type to correct for the lensing
effect of ``known'' halos by subtracting them from the complete convergence
maps.
The maps have been smoothed with a Gaussian with FWHM$=1'$ and
the scale is between -0.036 and 0.2.
}
\label{fig:NFW}
\end{figure}

\begin{figure}[h]
\begin{center}
\includegraphics[width=14cm]{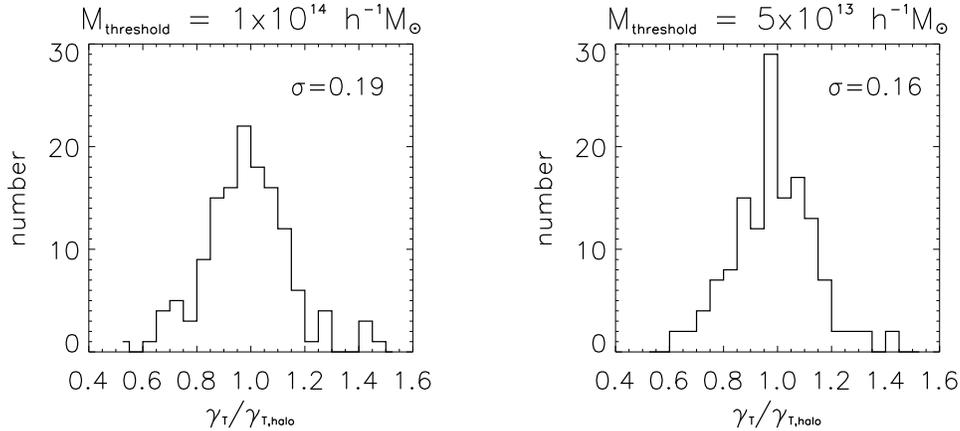}
\end{center}
\caption{Scatter in mass estimates based on the transverse shear for
a sample of 141 target halos defined in the text.
$\gamma_{\rm T}$
is the measured shear after correcting for the projection effect of line-of-sight halos more massive
than $M_{\rm threshold}$ assuming NFW profiles for those halos. $\gamma_{\rm T,halo}$ is
the shear purely caused by the halo of interest. In addition to the error caused by the projection
effect from halos below the thresholds, the slight extra scatter with respect to Fig. \ref{fig:hisreal} is caused
by the difference between true halo shapes and the assumed NFW model.
}
\label{fig:hisNFW}
\end{figure}

We also construct convergence maps using halo masses and positions from the halo catalog
assuming all halos follow NFW density profiles.
Earlier work shows that lensing maps using the NFW
profile can have power spectra that agree with power spectra obtained from simulated
matter distributions very well \citep{MaFry,Seljak}.
In Fig. \ref{fig:NFW} we show such a map 
only taking
into account halos more massive than $10^{14}\,h^{-1}\Msun$ next to the map
using the original simulation data with the same mass threshold. The two lensing fields
indeed
look fairly similar, even though it is obvious that the assumption that halos are
spherically symmetric is an idealization. We now use NFW convergence maps, only including
halos above certain mass thresholds,
to correct for the lensing effect of those halos
(Fig. \ref{fig:hisNFW}) by first subtracting
the NFW convergence map from the complete map and then calculating the tangential shear.
Comparison with the previous case where
we use the exact density profile to calculate the correction shows that assuming an NFW profile introduces
extra scatter.

The former results show
that our ability to retrieve $\gamma_{\rm T,halo}$ is severely limited by the lensing effect
of halos below the mass thresholds and mass that is not part of any halo at all. Only a small part of the scatter is
caused by the halos we correct for, defined by $M>M_{\rm threshold}$. This can be explained by the fact that each
target halo has only a small number
of these halos along its line of sight (Fig. \ref{fig:los}). We explicitly checked that as we go to
lower mass thresholds the histograms tighten significantly, but we must reduce $M_{\rm threshold}$
to values smaller than $10^{12}\,h^{-1}\Msun$ for the width to drop below 10 \%.
The NFW results, in addition, suffer a little from the fact
that we use idealized halo shapes to calculate the correction. We expect that this situation can be improved by
changing the halo model.
For example, one could allow for elliptical halos instead of only spherically symmetric ones and there may also
be some gain in
modelling halos beyond the virial radius. We will not pursue these approaches here though. As noted before, no matter
how much we improve our halo model, we will never be able to reduce the scatter further than in the case shown in
Fig. \ref{fig:hisreal} (unless we can find line-of-sight halos down to much lower masses).

\subsection{Mass based on cluster richness} \label{subsec:galcounts}

\begin{figure}[h]
\begin{center}
\includegraphics[width=14cm]{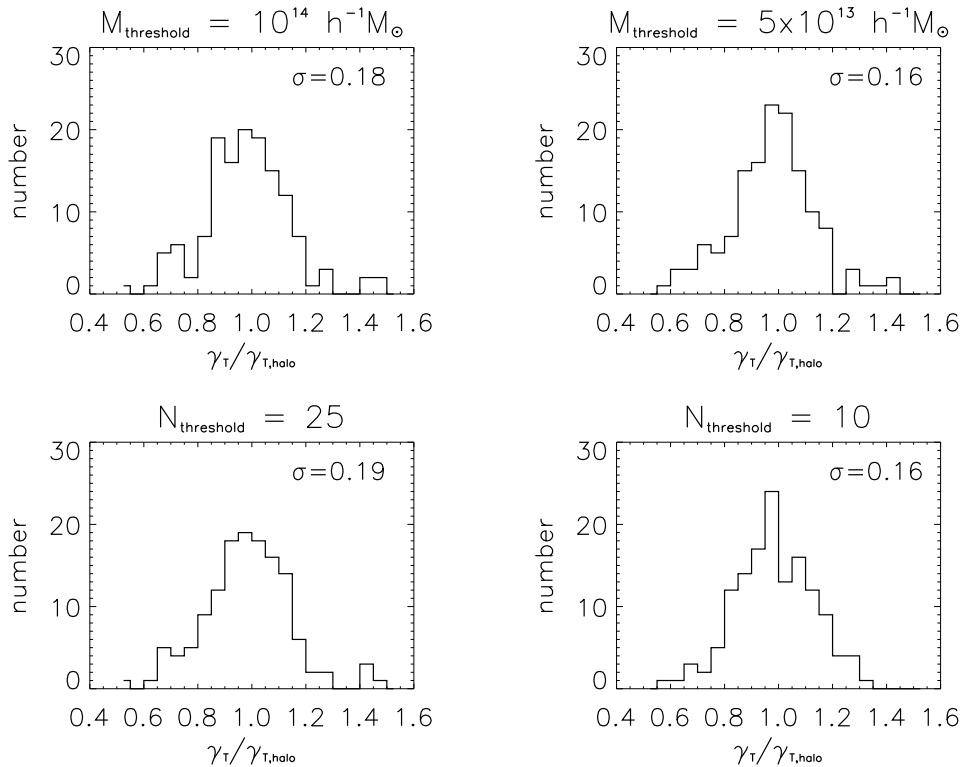}
\end{center}
\caption{Scatter in mass estimates based on the transverse shear for
a sample of 141 massive halos (see text).
$\gamma_{\rm T}$
is the measured shear after correcting for the projection effect of line-of-sight halos more massive
than $M_{\rm threshold}$ assuming NFW profiles and estimating their masses
from the cluster richness. We assume our survey detects all galaxies with magnitude below $25$.
$\gamma_{\rm T,halo}$ is
the shear purely caused by the halo of interest.
The upper plots assume all halos with $M>M_{\rm threshold}$ have
been identified and in the lower plots we correct for halos with the number of observed galaxies
$N>N_{\rm threshold}$.
}
\label{fig:richness}
\end{figure}

Instead of using the catalog masses for the line-of-sight halos it is
more realistic to use masses derived from an observable such as cluster
richness. We expect richness data will be the most readily available mass
estimator for a large sample of halos. Specifically we shall (optimistically)
assume all of the dark halos above certain threshold values of the number of
observed galaxies can be identified from an
optical survey. We use the number of galaxies as a mass estimator.
For a more detailed
account on galaxies in dark matter halos, we refer to the appendix. We first use
$\langle N \rangle = 30 (M/10^{15}\,h^{-1}\Msun)$
to calculate the expected number of
galaxies with $L > L_{\star}(z)$, where
$L_{\star}$ is defined by a corresponding absolute magnitude of
$M_{\star}(z)=-20.4-z$.
We then apply
a random Poisson scatter to the number of satellite galaxies to get a value of $N$
for each halo.
The number of observed
galaxies is defined to be the number of galaxies with magnitude smaller than
$25.0$ and is found using the luminosity function.
The line-of-sight halo masses we use to build our NFW maps are calculated
from this ``observed''
quantity by making the assumption $N = \langle N \rangle$ and inverting
the calculations described above. The resulting masses will deviate from
the real masses because of the Poisson scatter in $N-1$ and because of rounding.
However, one might expect this scatter not to be so important because the average
mass will not be affected and we are correcting for a number of line-of-sight halos.
Another reason why the mass scatter will likely not have a big effect is that, as we saw in
the previous section, the effect of
halos above the thresholds is very small anyway.

We now use these masses based on cluster richness and again the assumption of
NFW profiles to correct the tangential shear for projection effects.
The results are shown in Fig. \ref{fig:richness}
for different selections of halos. We first consider the case where all halos above
mass threshold of $1\times10^{14}$ and $5\times10^{13}\,h^{-1}\Msun$
have been identified to compare with results from
the previous section (top). As expected, the added uncertainty in mass does not
make a big difference. The most realistic cases we consider
are those corresponding to the results in the bottom two plots. Here, whether
a halo is considered to be observable or not is based on a minimum observable galaxy
number (10 and 25) instead of a minimum mass.

\section{Conclusions} \label{sec:conclusion}

Gravitational lensing provides a powerful means to measure the mass distribution in the
universe.
We were hoping that we could use the tangential shear caused by dark matter halos to find their masses.
A problem would be the effect of other mass along the line of sight on the shear. We investigated
if it is possible to use positions and masses of line-of-sight halos obtained from deep field optical data to correct for
this effect. In section \ref{subsec:findmass} we showed that
even if we can correct for the lensing effect of all line-of-sight halos with mass above
$5\times10^{13}\,h^{-1}\Msun$ perfectly, the effect of halos below that mass and unclustered matter is big enough to
make halo mass estimates based on the shear deviate strongly (20\% deviations are not very rare) from the actual
masses. Since in
reality we will not know the exact shapes of halos, this sets a limit on how accurately
we may hope to calculate individual halo masses.

This brings us to the second problem. In reality we will have to use a model for the shapes
of halos. Assuming all halos follow a NFW matter distribution causes estimated contributions
of line-of-sight halos to the lensing field to deviate from the actual contributions.
Of course, more realistic models could improve the results, but only up to
the limit mentioned above. Finally, in section \ref{subsec:galcounts} the use of galaxy counts
to find the masses of line-of-sight halos has been taken into account. We conclude that it is
very difficult to use the tangential shear to calculate masses of individual halos with high
accuracy and it is unlikely that any other measure of the lensing field {\em can} be used for this
purpose. An interesting future project would be to investigate the more statistical approach
of comparing predictions of the shear (or an other observable measure of the lensing field) as a function of halo mass
from simulations to measured lensing fields around a large sample of halos
to find the number of halos per mass range. While it has never been demonstrated that high accuracy
calibration of the shear-mass relation is possible using simulations, there is in principle no
obstacle to this route. Of course, using simulations to calibrate observations, rather than test
algorithms, is more demanding of the theory.

We would like to thank Chris Vale for useful discussions and also Alexandre Amblard and Joseph
Hennawi for their helpful comments on an earlier draft.

\section{Appendix: Galaxy counts}\label{sec:app}

Generally, a halo contains $1$ central galaxy and $N-1$ satellite galaxies above a certain
reference luminosity $L_{\star}$, where
the number of satellite galaxies follows a Poisson
distribution \citep{Kravtsov}. The mean number of galaxies as a function of mass is given by
the relation $\langle N \rangle \sim M^{\alpha}$, with $\alpha \approx 0.9$. The number
of galaxies that we will be able to observe will likely be determined by an upper
limit on the apparent magnitude, which can be related to a minimum luminosity $L_{\rm min}$
if we know the physical distance to the galaxy/halo. The number of galaxies per volume in
a given luminosity range is given by the Schechter luminosity function:

\begin{equation}
\phi(L)dL = \phi_{\star}\left[\frac{L}{L_{\star}}\right]^{\beta}e^{-L/L_{\star}}dL \ \ \ \  (\beta \approx -1),
\end{equation}

from which it follows that

\begin{equation}
N_{L>L_{\rm min}} = N \,\frac{\Gamma(\beta+1,L_{min}/L_{\star})}{\Gamma(\beta+1,1)},
\end{equation} 

where $N_{L>L_{min}}$ is the number of galaxies in the halo with $L > L_{min}$,
$N$ is the number of galaxies with $L > L_{\star}$ discussed above and
$\Gamma(\alpha,x)$ is the incomplete gamma function. The relations above can be used
to relate the number of observed galaxies to the halo mass. Of course, the accuracy
of this method is limited because of the Poisson scatter in the number of satellite galaxies
and because the relations used are far from exact.

\end{document}